\documentclass[
    ,final            % use final for the camera ready runs
  ]{aipproc}

\layoutstyle{6x9}

\graphicspath{{./Figs/}}
\usepackage{epsfig}

\newcommand{\preprint}{
  \begin{picture}(0,0)
  \put(346,325){\rm\small HU--EP--06/39}
  \put(325,305){\rm\small ITEP-LAT/2006-09}
  \end{picture}}

\newcommand{\ie}{{\it i.e. }}
\newcommand{\noi}{\noindent}
\newcommand{\beq}{\begin{equation}}
\newcommand{\eeq}{\end{equation}}
\newcommand{\bea}{\begin{eqnarray}}
\newcommand{\eea}{\end{eqnarray}}

\begin{document}

\title{Topology and confinement at $T \neq 0$: \\
calorons with non-trivial holonomy\footnote{Talk by M. M\"uller-Preussker 
at {\it Quark Confinement and Hadron Structure VII}, Ponta Delgada, Azores, 
Portugal, September 2 - 7, 2006
}}

\author{P. Gerhold}{
address={Humboldt-Universit\"at zu Berlin, Institut f\"ur Physik,
Newtonstr. 15, 12489 Berlin, Germany}}
\author{E.-M. Ilgenfritz}{
address={Humboldt-Universit\"at zu Berlin, Institut f\"ur Physik,
Newtonstr. 15, 12489 Berlin, Germany}}
\author{M. M\"uller-Preussker}{
address={Humboldt-Universit\"at zu Berlin, Institut f\"ur Physik,
Newtonstr. 15, 12489 Berlin, Germany}}
\author{B.V. Martemyanov}{
address={Institute for Theoretical and Experimental Physics,
B. Cheremushkinskaya 25, Moscow 117259, Russia}}
\author{A.I. Veselov}{
address={Institute for Theoretical and Experimental Physics,
B. Cheremushkinskaya 25, Moscow 117259, Russia}}

\begin{abstract}
In this talk, relying on experience with various lattice filter 
techniques, we argue that the semiclassical structure of finite 
temperature gauge fields for $T < T_c$ is dominated by calorons 
with non-trivial holonomy. By simulating a dilute gas of calorons 
with identical holonomy, superposed in the algebraic gauge,
we are able to reproduce the confining properties below $T_c$ up
to distances $r=O(4 \mathrm{fm}) >> \rho$ (the caloron size).
We compute Polyakov loop correlators as well as space-like 
Wilson loops for the fundamental and adjoint representation. 
The model parameters, including the holonomy, can be inferred
from lattice results as functions of the temperature. 
\end{abstract}

\date{\today}

\maketitle

\preprint

\vspace*{-0.6cm}
Instanton or caloron models of QCD successfully describe many
non-perturbative features of hadron physics, in particular 
chiral symmetry breaking and the $U_A(1)$ anomaly 
(for reviews see~\cite{Schafer:1996wv,Diakonov:2002fq}). 
However, they fail to describe confinement unless they are 
endowed with long-range correlations. 
This is the case for instantons in the regular gauge~\cite{Negele:2004hs}
and has been discussed also at this conference~\cite{Wagner:2006qn}. 
An attractive alternative, at least for non-zero temperature $T$, 
is based on new caloron solutions with non-trivial 
holonomy~\cite{Kraan:1998kp,Kraan:1998pm,Kraan:1998sn,Lee:1998bb},
worked out in various aspects by Kraan and van Baal. 
At this symposium we were happy to listen to a review 
talk~\cite{vanBaal:2006nx} by Pierre van Baal after his recovery.

The new caloron solutions - we call them KvBLL calorons - have 
characteristic properties which distinguish them from the old 
BPST instantons~\cite{Belavin:1975fg} or Harrington-Shepard (HS) 
calorons~\cite{Harrington:1978ve}. Like the latter, the new calorons are 
(anti)selfdual with integer topological charge and periodic in 
Euclidean time with the period $1/T$.
The difference is a non-trivial asymptotic behaviour of $A_4(x)$
such that the Polyakov loop 
\beq
P(\vec x) = \hat P\, \exp\left( i \int\limits_0^{1/T} A_4(\vec x,t) dt\right) 
~~\stackrel{|\vec{x}| \to \infty}{\Longrightarrow}~~
{\cal P}_{\infty} 
\label{eq:holonomy}
\eeq
can take arbitrary fixed values ${\cal P}_{\infty} \notin Z(N_c)$ 
at spatial infinity.
Each single KvBLL (anti)caloron consists of $N_c$ monopole constituents
localized at positions where the Polyakov loop $P(\vec{x})$ has 
degenerated eigenvalues. For $SU(2)$ this means that 
$L(\vec{x}) = \frac{1}{2} \mathrm{tr} P(\vec{x})$ takes opposite values 
$\pm 1$ at the positions of the two monopoles. The profile of the Polyakov 
loop field inside a KvBLL solution is the most significant feature of 
the new calorons irrespective whether the constituents or 'instanton quarks' 
are separated or not. If the constituents are far from each other the 
caloron dissociates into $N_c$ static non-Abelian monopoles. 
The topological charge of either lump then depends on the eigenvalue 
differences of ${\cal P}_{\infty}$. The zero modes of the Dirac operator 
are localized only at one of the constituents.  
When the fermionic boundary condition is smoothly changed the 
zero mode jumps from one constituent to another~\cite{Chernodub:1999wg} 
if these are separated. 

First we used the cooling method applying it to pure $SU(2)$ and $SU(3)$ 
lattice Monte Carlo gauge fields. We demonstrated that the lumps of 
topological charge observed in the plateau configurations have to be 
interpreted in terms of KvBLL calorons
~\cite{Ilgenfritz:2002qs,Ilgenfritz:2005um}. Closer to the deconfinement
transition or for a smaller aspect ratio we found an increasing frequency
of dissociated monopole pairs in $SU(2)$.  

\begin{figure}
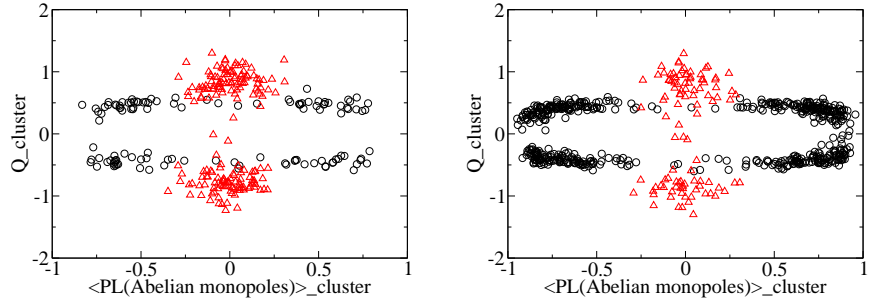

\includegraphics[height=.18\textheight]{plqcorrclust23new.eps}
\hspace{0.5 cm}
\includegraphics[height=.18\textheight]{plqcorrclust24.eps} 
\caption{Scatter plots of topological charge versus averaged Polyakov loop
for topological clusters of lattice fields produced with the
Wilson action at $\beta=2.3,2.4$ and lattice size
$24^3 \times 6$ (confinement). Triangles (circles) denote full caloron 
(isolated static monopole) candidates.
}
\label{fig:cluster_conf}
\end{figure}

More recently we have studied Monte Carlo lattice fields with the 4d smearing 
method at different temperatures~\cite{Ilgenfritz:2006ju}.
We found many clusters of topological charge and classified them with 
respect to their Abelian monopole content (in the maximally Abelian gauge).
Two limiting cases suggest an interpretation in terms of KvBLL constituents 
or calorons: 
(i) clusters containing a monopole loop winding around the lattice in the
time direction, taken as candidates for a single constituent;
(ii) clusters containing a closed monopole loop, taken as candidates 
for undissociated calorons. 
For these cases we have estimated the topological charge of the cluster, 
$Q_{\rm cluster}$, and the Polyakov loop averaged over the positions 
of time-like Abelian monopoles, $<PL>_{\rm cluster}$.
In the confinement phase - \ie for maximally non-trivial holonomy - 
we would expect half-integer topological charges for isolated monopoles 
and integer charge for full calorons. The averaged Polyakov loop 
should be close to $\pm 1$ for isolated monopoles 
and near zero for calorons according to the ``dipole'' profile
of the Polyakov loop inside the KvBLL caloron. 
What is really observed is seen in the scatter plots in the
$(Q_{\rm cluster},<PL>_{\rm cluster})$-plane of Fig.~\ref{fig:cluster_conf}. 
Each entry corresponds to one of the selected cluster candidates,
and the scatter plot is clustering into classes with the expected signatures.
In the deconfined phase (not shown), for holonomies closer to the trivial 
one we would expect to find disbalanced constituents, one with small 
action and a complementary one with large action.
Apart from few full calorons accounting for the topological charge  
of the configurations, we found many ``single-constituent'' clusters 
with static Abelian monopole loops but small topological charge, whereas 
constituents with topological charges close to $\pm 1$ were completely missing. 
We conclude that the model picture of KvBLL calorons may fail in the 
deconfinement phase.   
%----------------------------------------------------------------------------
%----------------------------------------------------------------------------

One can also study the topological content without cooling or smearing 
techniques by applying purely fermionic methods to equilibrium fields.
Such investigations have also provided indications for the presence
of KvBLL monopole constituents~\cite{Gattringer:2002tg,Gattringer:2004pb}.   

What are the consequences if HS calorons are replaced by KvBLL ones
as building blocks in a random caloron gas model at finite $T$?
Such a model, so far realized in the $SU(2)$ case 
\cite{Gerhold:2006sk}, requires to start the superposition in 
the so-called algebraic gauge for which $A_4(x)$
decreases sufficiently fast. A non-periodic gauge transformation is applied 
in order to render the gauge field periodic.
This restricts us to superpositions of calorons with identical holonomy.
The positions of the calorons are chosen randomly, the sizes $\rho$ ({\it i.e.}
the distances $d=\pi \rho^2 / \beta$ between the constituents) are sampled
for $T>T_c$ according to~\cite{Gross:1980br}
\beq
D(\rho,T) = 
A(T) \cdot \rho^{b-5}\cdot \exp(-\frac{4}{3}(\pi T \rho)^2)\,,
~~~~b=11 N_c/3=22/3\,. 
\eeq
For $T<T_c$ temperature independence is postulated but keeping the 
suppression at $T_c$ fixed (see \cite{Ilgenfritz:1980vj,Diakonov:1983hh}).
For a statistically uncorrelated caloron gas the actual density 
$n(T)$ can be inferred from lattice computations of the topological 
susceptibility. The average size was fixed by comparison between
model and lattice results for the spatial string tension in units of the 
critical temperature which then turned out to be $T_c \simeq 178 \mathrm{MeV}$.  
The holonomy  ${\cal P}_{\infty} \equiv \exp (2\pi i \omega \tau_3)$  
was identified with the (renormalized) Polyakov loop. 
More details and references can be found in~\cite{Gerhold:2006sk}. 
For some parameter sets see Table \ref{tab:parameters}.    
%-----------------------------------------------------------------------------
\begin{table}
\begin{tabular}{|c|c||c|c|c||c||c|c|c|}
\hline
$T/T_c$ & $N_s^3\times N_{\tau}$ & $n^{\frac{1}{4}}\,[{\rm MeV}]$ & $4\omega$ 
&  $\bar\rho\,[{\rm fm}]$& \# & $cos(2\pi\omega)$& $<|L|>$ 
&  $\gamma$  \\ 
\hline
$0.80$  & $32^3\times 10$& 198 & 1.00 & $ 0.37 $   
		& 777 & 0.00  & $0.13(1)$ & $1.61(1)$ \\
%$0.90$  & $32^3\times 9$ & 198 & 1.00 & $ 0.37 $   
%		& 591 & 0.00  & $0.14(1)$ & $1.65(1)$ \\
$1.00$  & $32^3\times 8$ & 198 & 1.00 & $ 0.37 $   
		& 526 & 0.00  & $0.14(1)$ & $1.69(1)$ \\ 
%$1.10$  & $32^3\times 8$ & 178 & 0.61 & $ 0.33 $   
%		& 160 & 0.58  & $0.43(1)$ & $1.29(1)$ \\
$1.20$  & $32^3\times 8$ & 174 & 0.51 & $ 0.31 $   
		& 160 & 0.70  & $0.59(1)$ & $1.18(1)$ \\
%$1.32$  & $32^3\times 8$ & 165 & 0.43 & $ 0.28 $   
%		& 160 & 0.78  & $0.72(1)$ & $1.10(1)$ \\
\hline
\end{tabular}
\caption{Model parameters $n(T),\,\omega(T),\,\bar\rho(T)$, the
lattice grid size ${N_s \times N_{\tau}}$, and the number of generated 
configurations \#  for selected temperature values
$T/T_c$. Furthermore, the measured average Polyakov loop 
$<|L|>$ (together with the input value ${cos(2\pi\omega)}$) 
and the action surplus factor 
$\gamma = \frac{S_{tot}}{N_{caloron} \cdot S_{inst}}$ are given.}
\label{tab:parameters}
\end{table}
%----------------------------------------------------------------------------

\begin{figure}
\includegraphics[height=0.20\textheight]{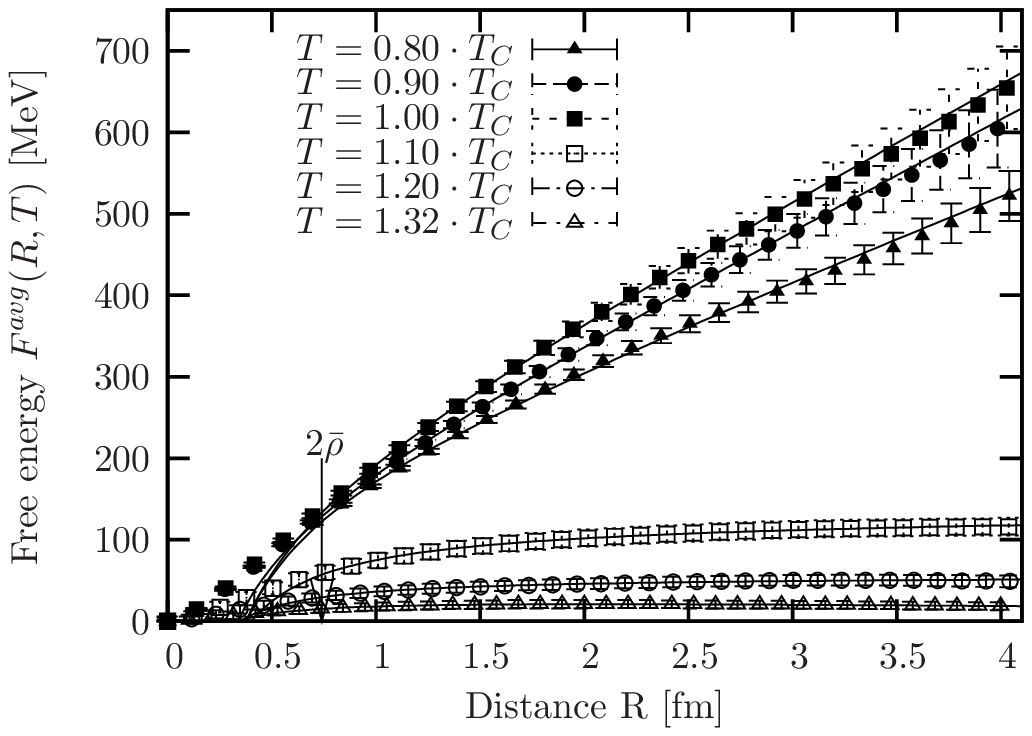}   
\hspace*{0.5cm}  
\includegraphics[height=0.20\textheight]{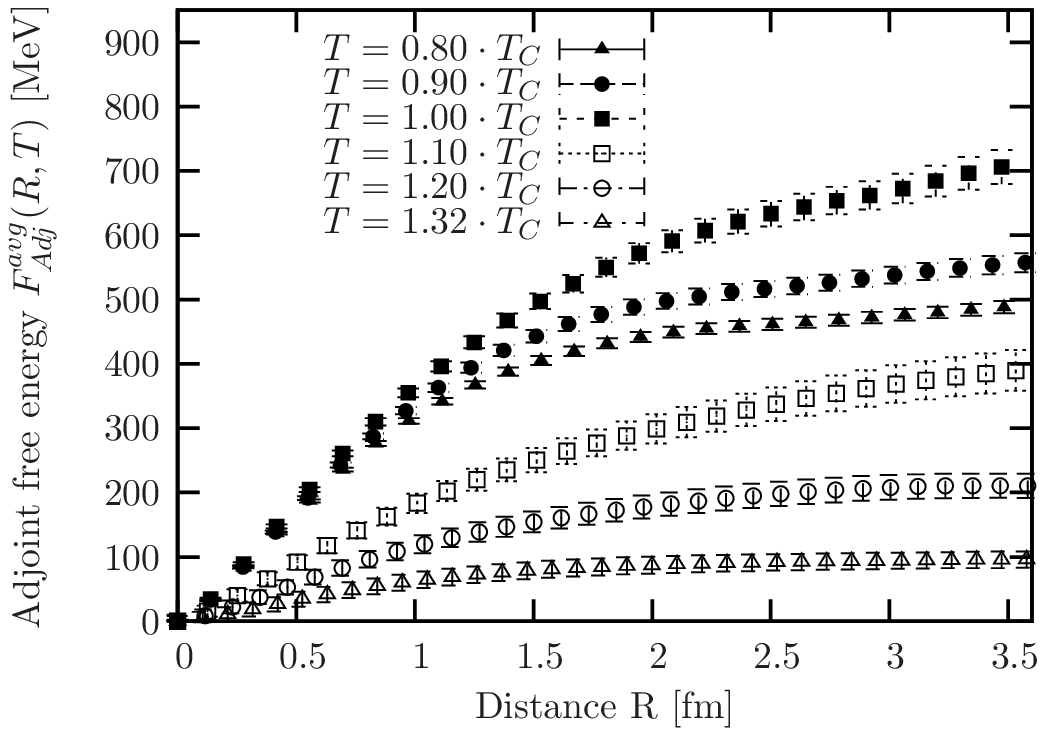}  
\caption{Colour averaged free energy versus distance $R$ at various 
temperatures for the fundamental (left) and adjoint (right) representations.}
\label{fig:free_energy}
\end{figure}

On a lattice grid we have computed spatial Wilson loops as well as
Polyakov loop correlators, both within fundamental and adjoint
representations. The spatial string tension is seen to drop at $T_c$, 
because the mechanism responsible for the observed rise are not the
monopoles that are part of the calorons and suppressed at $T>T_c$.
This problem corresponds to our observation, in the smeared configurations 
reported above, of many monopoles with low accompanying topological charge.
The results for the free energy of a static quark-antiquark 
pair are shown in Fig. \ref{fig:free_energy}. 
We can follow a linear rise for distances up to $O(4 \mathrm{fm})$ in 
the confinement phase, whereas it becomes screened above $T_c$. 
The free energy of adjoint charge pairs is screened in both phases.

%----------------------------------------------------------------------------
%----------------------------------------------------------------------------
We conclude that a semiclassical model for finite-temperature $SU(2)$ fields 
should start from calorons with generic holonomy. 
Such a model turns out to describe confinement with parameters which are
rather close to standard instanton model assumptions. Keeping all parameters 
fixed, merely changing the holonomy from $\omega=1/4$ to $\omega=0$ or $1/2$,
removes the linear rise completely \cite{Gerhold:2006sk}. 
This underscores the r\^ole of non-trivial holonomy and the corresponding 
long-range nature of the caloron fields.

\vspace*{0.3cm}\noi
We are grateful to P. van Baal, F. Bruckmann, C. Gattringer, A. Sch\"afer 
and S. Solbrig for numerous useful discussions. We acknowledge financial 
support by the DFG through FOR 465 / Mu 932/2-4 and 436 RUS 113/739/0-2,
by RFBR through grants 04-02-16079 and 06-02-16309 as well as by the 
RFBR-DFG grant 06-02-04010.

\vspace*{-0.5cm}
\bibliographystyle{aipproc}   % if natbib is available
\bibliography{caloron}

\begin{thebibliography}{21}
\expandafter\ifx\csname natexlab\endcsname\relax\def\natexlab#1{#1}\fi
\providecommand{\enquote}[1]{``#1''}
\expandafter\ifx\csname url\endcsname\relax
  \def\url#1{\texttt{#1}}\fi
\expandafter\ifx\csname urlprefix\endcsname\relax\def\urlprefix{URL }\fi
\providecommand{\eprint}[2][]{\url{#2}}

\bibitem[Sch{\"a}fer and Shuryak(1998)]{Schafer:1996wv}
T.~Sch{\"a}fer, and E.~V. Shuryak, \emph{Rev. Mod. Phys.} \textbf{70}, 323
  (1998), \eprint{hep-ph/9610451}.

\bibitem[Diakonov(2003)]{Diakonov:2002fq}
D.~Diakonov, \emph{Prog. Part. Nucl. Phys.} \textbf{51}, 173 (2003),
  \eprint{hep-ph/0212026}.

\bibitem[Negele et~al.(2005)]{Negele:2004hs}
J.~W. Negele, F.~Lenz, and M.~Thies, \emph{Nucl. Phys. Proc. Suppl.}
  \textbf{140}, 629 (2005), \eprint{hep-lat/0409083}.

\bibitem[Wagner(2006)]{Wagner:2006qn}
M.~Wagner  (2006), \eprint{hep-ph/0608090}.

\bibitem[Kraan and van Baal(1998{\natexlab{a}})]{Kraan:1998kp}
T.~C. Kraan, and P.~van Baal, \emph{Phys. Lett.} \textbf{B428}, 268
  (1998{\natexlab{a}}), \eprint{hep-th/9802049}.

\bibitem[Kraan and van Baal(1998{\natexlab{b}})]{Kraan:1998pm}
T.~C. Kraan, and P.~van Baal, \emph{Nucl. Phys.} \textbf{B533}, 627
  (1998{\natexlab{b}}), \eprint{hep-th/9805168}.

\bibitem[Kraan and van Baal(1998{\natexlab{c}})]{Kraan:1998sn}
T.~C. Kraan, and P.~van Baal, \emph{Phys. Lett.} \textbf{B435}, 389--395
  (1998{\natexlab{c}}), \eprint{hep-th/9806034}.

\bibitem[Lee and Lu(1998)]{Lee:1998bb}
K.-M. Lee, and C.-H. Lu, \emph{Phys. Rev.} \textbf{D58}, 025011 (1998),
  \eprint{hep-th/9802108}.

\bibitem[van Baal(2006)]{vanBaal:2006nx}
P.~van Baal  (2006), \eprint{hep-ph/0610409}.

\bibitem[Belavin et~al.(1975)]{Belavin:1975fg}
A.~A. Belavin, A.~M. Polyakov, A.~S. Shvarts, and Y.~S. Tyupkin, \emph{Phys.
  Lett.} \textbf{B59}, 85 (1975).

\bibitem[Harrington and Shepard(1978)]{Harrington:1978ve}
B.~J. Harrington, and H.~K. Shepard, \emph{Phys. Rev.} \textbf{D17}, 2122
  (1978).

\bibitem[Chernodub et~al.(2000)]{Chernodub:1999wg}
M.~N. Chernodub, T.~C. Kraan, and P.~van Baal, \emph{Nucl. Phys. Proc. Suppl.}
  \textbf{83}, 556--558 (2000), \eprint{hep-lat/9907001}.

\bibitem[Ilgenfritz et~al.(2002)]{Ilgenfritz:2002qs}
E.-M. Ilgenfritz, B.~V. Martemyanov, M.~M{\"u}ller-Preussker, S.~Shcheredin,
  and A.~I. Veselov, \emph{Phys. Rev.} \textbf{D66}, 074503 (2002),
  \eprint{hep-lat/0206004}.

\bibitem[Ilgenfritz et~al.(2005)]{Ilgenfritz:2005um}
E.~M. Ilgenfritz, M.~Muller-Preussker, and D.~Peschka, \emph{Phys. Rev.}
  \textbf{D71}, 116003 (2005), \eprint{hep-lat/0503020}.

\bibitem[Ilgenfritz et~al.(2006)]{Ilgenfritz:2006ju}
E.~M. Ilgenfritz, B.~V. Martemyanov, M.~M{\"u}ller-Preussker, and A.~I.
  Veselov, \emph{Phys. Rev.} \textbf{D73}, 094509 (2006),
  \eprint{hep-lat/0602002}.

\bibitem[Gattringer and Sch{\"a}fer(2003)]{Gattringer:2002tg}
C.~Gattringer, and S.~Sch{\"a}fer, \emph{Nucl. Phys.} \textbf{B654}, 30 (2003),
  \eprint{hep-lat/0212029}.

\bibitem[Gattringer and Pullirsch(2004)]{Gattringer:2004pb}
C.~Gattringer, and R.~Pullirsch, \emph{Phys. Rev.} \textbf{D69}, 094510 (2004),
  \eprint{hep-lat/0402008}.

\bibitem[Gerhold et~al.(2006)]{Gerhold:2006sk}
P.~Gerhold, E.~M. Ilgenfritz, and M.~M{\"u}ller-Preussker  (2006),
  \eprint{hep-ph/0607315}.

\bibitem[Gross et~al.(1981)]{Gross:1980br}
D.~J. Gross, R.~D. Pisarski, and L.~G. Yaffe, \emph{Rev. Mod. Phys.}
  \textbf{53}, 43 (1981).

\bibitem[Ilgenfritz and M{\"u}ller-Preussker(1981)]{Ilgenfritz:1980vj}
E.-M. Ilgenfritz, and M.~M{\"u}ller-Preussker, \emph{Nucl. Phys.}
  \textbf{B184}, 443 (1981).

\bibitem[Diakonov and Petrov(1984)]{Diakonov:1983hh}
D.~Diakonov, and V.~Y. Petrov, \emph{Nucl. Phys.} \textbf{B245}, 259 (1984).

\end{thebibliography}

\end{document}